# Segregation-Controlled Diffusion-Induced Grain Boundary Migration in Alloy 690

Yuntian Jiang[1], Shuai Zhang[1], Zhuoming Xie[2], Xiaohan Bie[3], Huiqiu Deng[4], Wangyu Hu[1], Jie Hou[1*]

1. College of Materials Science and Engineering, State Key Laboratory of Cemented Carbide, Hunan University, Changsha 410082, China
2. Key Laboratory of Materials Physics, Institute of Solid State Physics, Chinese Academy of Sciences, P. O. Box 1129, Hefei 230031, PR China
3. Department of Mining and Materials Engineering, McGill University, Montreal, Quebec, H3A 0C5, Canada
4. School of Physics and Electronics, Hunan University, Changsha 410082, China

\* Corresponding author: jiehou@hnu.edu.cn (Jie Hou)

## Abstract

Grain boundary (GB) migration accompanied by Cr depletion is widely observed in Alloy 690 and is closely linked to intergranular degradation and stress corrosion cracking. However, the fundamental driving force for GB migration and its link with Cr depletion remains unclear. In this work, hybrid molecular dynamics and semi-grand canonical Monte Carlo simulations were employed to investigate GB migration in Alloy 690 under coupled solute diffusion and segregation effects across a range of GB characters. The results show that Cr segregation at GBs, while generally considered favorable for GB stability, can facilitate diffusion-induced GB migration and Cr depletion. Cr diffusion along GBs produces localized Cr depletion zones that are energetically incompatible with positively segregating GBs, generating a chemical driving force that drives GB migration toward the Cr-rich matrix, which ultimately results in persistent GB migration accompanied by a Cr depletion. By quantifying solute–GB interaction energetics, we demonstrate that GB migration is quantitively controlled by the coupled effects of solute diffusivity and segregation strength. These mechanistic insights provide a unified framework that rationalizes experimentally observed correlations between GB character, Cr depletion, and GB migration in Cr-containing alloys.

## 1.Introduction

Metals and their alloys play indispensable roles in modern infrastructure owing to their favorable mechanical properties and high electrical and thermal conductivities. However, corrosion remains the dominant degradation mechanism limiting their long-term reliability, leading not only to performance deterioration and shortened service life but also to severe safety risks and substantial economic losses. It has been widely reported that corrosion-related damage accounts for approximately 3%–5% of GDPs worldwide[1, 2], underscoring the critical importance of understanding and mitigating corrosion processes.

In extreme service environments such as pressurized water reactors (PWRs), structural components are subjected simultaneously to high-temperature aqueous corrosive, complex mechanical stresses, and neutron irradiation, rendering corrosion control far more challenging than in conventional conditions[3-5]. Cr is therefore widely alloyed into commercial Ni- and Fe-based alloys to enhance corrosion and oxidation resistance through formation of a protective Cr-rich surface oxide[6-9]. For instance, alloy 690 contains approximately twice the Cr content of alloy 600 (~30 wt.% vs. ~15 wt.%), exhibits markedly improved resistance to stress corrosion cracking (SCC) initiation and has consequently become the preferred material for key PWR components[10-13]. Nevertheless, crack propagation has still been observed in pre-cracked Alloy 690 specimens exposed to hydrogenated steam environments[14-20]. These observations indicate that while Cr effectively reduces crack initiation, it does not fully eliminate SCC crack propagation, highlighting an unresolved mechanistic issue in the SCC behavior of Cr-containing alloys.

Grain boundaries (GBs) play a central role in the SCC behavior of Ni-based alloys. As established by research, the degradation of alloys is, in essence, an intergranular phenomenon[20-25]. It's generally believed that initiation of intergranular SCC in Alloy 690 (and other Cr-containing alloys) starts with Cr diffusion along GBs in oxidizing environments to form a protective $Cr_2O_3$ film. This film ruptures and re-passivates repeatedly under stress, leading to preferential intergranular oxidation and continuous Cr consumption. As oxidation proceeds, a Cr depletion zone develops beneath the crack tip along the GB. Once the local Cr concentration (typically reduced to 10%–20 wt.%) becomes insufficient to sustain a protective chromia layer, crack nucleation and propagation occur. Increasing evidence suggests that the SCC susceptibility is closely associated with its tendency for intergranular oxidation driven by localized Cr depletion. Consequently, the formation of Cr depletion zones is widely regarded as a characteristic "precursor event" during the early stages of pressurized water SCC in Ni-based alloys[26].

Recent studies have revealed a strong correlation between Cr depletion and GB migration. For instance, Kuang et al.[8] reported that the Cr depletion varies significantly among different GB types, with random high angle boundary exhibit pronounced migration accompanied by wide Cr depletion zones, whereas incoherent twin boundary display a narrow Cr depletion zone without apparent migration, and coherent twin boundary show negligible migration and Cr depletion. Feng et al. [27]further link the tendency for passive-film formation and intergranular Cr depletion with the Cr diffusivity along GBs, showing that GBs with higher atomic packing density show slower Cr diffusion, shallower GB migration zones, and stronger resistance to intergranular oxidation. These observations indicate that diffusion kinetics governs GB migration behavior, and the associated GB migration is therefore commonly interpreted within the framework of diffusion-induced grain boundary migration (DIGM)[6, 22, 24, 25, 28].

Although DIGM provides a useful phenomenological description of diffusion-associated GB migration, increasing evidence suggests that the interaction between solute and GBs also plays a critical role in governing the driving force for GB migration, Brokman et al. [29]proposed that segregation-induced concentration differences across the GBs generate the chemical potential gradients necessary for migration. Moreover, research has further indicated that DIGM occurs only when solute segregation at GBs exceeds a certain threshold[30-34], while Kaur et al. [35] quantitatively linked solute concentration to the driving force for GB migration by distinguishing between segregation and anti-segregation behaviors. Notably, while the importance of solute–grain boundary interactions has been increasingly recognized, existing studies remain largely scattered and fragmented. As a result, a coherent and unified mechanistic framework that explicitly captures the coupled roles of solute diffusion and segregation in driving grain boundary migration is still lacking.

To address these issues, in this work, hybrid molecular dynamic and semi-grand canonical Monte Carlo (MD/SGCMC) simulations were employed. We systematically investigated GB migration in Alloy 690 under the coupled effects of solute diffusion and segregation for different types of GBs. By directly comparing Cr- and Fe-dominated diffusion systems, we isolate the solute-specific contributions to DIGM. The results reveal that Cr diffusion initiates the formation of localized Cr depletion at the GB, thereby generating a concentration gradient relative to the surrounding matrix. Owing to the positive segregation tendency of Cr toward GBs, this gradient produces a persistent attractive force that drives GB migration toward the Cr-rich matrix. As a result, a seemingly counterintuitive phenomenon is observed, whereby Cr segregation ultimately promotes Cr depletion along migrating GBs. Furthermore, the driving force for GB migration is quantitatively evaluated in terms of solute segregation strength, which, when combined with solute diffusivity, enables quantitative prediction of GB migration rates. These findings demonstrate that solute segregation plays a critical role in diffusion-induced grain boundary migration, providing new mechanistic insight into GB migration and Cr depletion in Cr-containing alloys.

## 2. Computational methods

To investigate DIGM in the Ni–Cr–Fe system, atomistic models were constructed based on two ternary alloys, denoted as Type-690 and Type-690(Fe). The Type-690 model represents the typical composition of commercial Alloy 690, containing 60 at.% Ni, 30 at.% Cr, and 10 at.% Fe, while the Type-690(Fe) model serves as a control system in which the atomic fractions of Cr and Fe are interchanged (60 at.% Ni, 30 at.% Fe, and 10 at.% Cr) to isolate the role of Cr in regulating GB migration behavior. Both models adopt a face-centered cubic (fcc) crystal structure, with atomic species randomly assigned to lattice sites.

The overall atomic configuration is illustrated schematically in Fig. 1. A symmetric tilt GB bi-crystal was constructed by rotating two grains in opposite directions by the same misorientation angle, resulting in symmetric tilt GB configurations. As detailed in Table 1, twelve representative GBs were examined, covering a broad range of crystallographic characters. A bulk single-crystal model was also simulated as a reference system. To introduce a controlled solute concentration gradient, a solute-depleted layer was constructed above the GB by selectively reducing the local concentration of Cr (for

Type-690) or Fe (for Type-690(Fe)) from the bulk value (30 at.%) to zero. This artificial depletion zone establishes a chemical potential gradient that drives solute diffusion along the GB, enabling direct simulation of solute-depletion-driven GB migration phenomena. Periodic boundary conditions were applied in all three spatial directions. The simulation box size is approximately $13 \times 30 \times 25\ nm^3$. with its precise dimensions varying depending on the periodicity of the specific grain boundary.

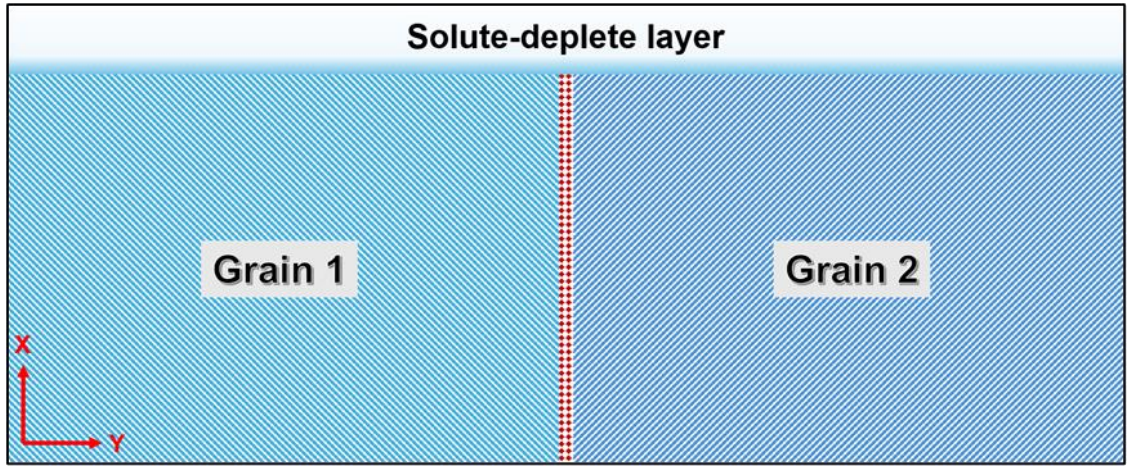

Figure 1. Schematic diagram of the bi-crystal model used in this work. Solute-deplete layer represent a region where Cr (or Fe) solutes are replaced by Ni atoms with SGCMC algorithm.

The coupled effects of solute diffusion and segregation were simulated using a hybrid MD/SGCMC framework implemented in the Large-scale Atomic/Molecular Massively Parallel Simulator (LAMMPS)[36]. In this approach, atomic positions were evolved using standard MD to capture structural relaxation and GB migration, while compositional redistribution was sampled through SGCMC species-exchange moves between Ni-Cr or Ni-Fe species. These exchanges were accepted according to the Metropolis criterion with a very low chemical potential for Cr or Fe to ensure zero solute concentration in the depleted layer. All MD simulations were carried out at 1400 K under an isothermal–isobaric (NPT) ensemble with a timestep of 1 fs. Each system was first minimized using the conjugate-gradient method, then equilibrated for 100,000 MD steps, followed by MD/SGCMC simulations during which one SGCMC exchange step was performed after every 20 MD steps. Atomic interactions in the Ni–Cr–Fe system were described using an embedded-atom method (EAM) potential developed by Daramola et al.[37], which accurately captures GB energetics and solute segregation behavior and ensures the reliability of the simulation result. Structural evolution and defect configurations were analyzed using OVITO[38].

Density functional theory (DFT) calculations were performed using VASP to benchmark the segregation energies of Cr and Fe at the Σ5 (3 1 0) GB[39, 40]. The PAW method and the GGA-PBE functional were employed, and spin polarization was explicitly included in all calculations[41, 42]. A GB supercell with size of $11.04 \times 21.82 \times 3.5$ Å$^3$ was used with a plane-wave cutoff energy of 350 eV and a gamma-centered Monkhorst–Pack k-point mesh of $3 \times 1 \times 9$. Structural relaxations were carried out until the total energy and atomic forces converged to $10^{-6}$ eV and 0.01 eV/Å, respectively[39].

Table 1. Tilt angles and crystallographic parameters of the twelve representative GBs investigated in this study.

| GB type | Tilt angle (°) | Upper grain | | | Lower grain | | |
|---|---|---|---|---|---|---|---|
| | | X | Y | Z | X | Y | Z |
| Σ3(111) | 70.53 | [$\bar{1}$12] | [1$\bar{1}$1] | [110] | [$\bar{1}$1$\bar{2}$] | [$\bar{1}$11] | [110] |
| Σ9(221) | 38.94 | [$\bar{1}$14] | [2$\bar{2}$1] | [110] | [$\bar{1}$1$\bar{4}$] | [$\bar{2}$21] | [110] |
| Σ19(331) | 26.5 | [$\bar{1}$16] | [3$\bar{3}$1] | [110] | [$\bar{1}$1$\bar{6}$] | [$\bar{3}$31] | [110] |
| Σ33(441) | 159.95 | [$\bar{1}$18] | [4$\bar{4}$1] | [110] | [$\bar{1}$1$\bar{8}$] | [$\bar{4}$41] | [110] |
| Σ27(552) | 148.41 | [$\bar{1}$15] | [5$\bar{5}$2] | [110] | [$\bar{1}$1$\bar{5}$] | [$\bar{5}$52] | [110] |
| Σ51(772) | 157.16 | [$\bar{1}$17] | [7$\bar{7}$2] | [110] | [$\bar{1}$1$\bar{7}$] | [$\bar{7}$72] | [110] |
| Σ5(310) | 36.87 | [03$\bar{1}$] | [013] | [100] | [031] | [0$\bar{1}$3] | [100] |
| Σ13(320) | 67.38 | [03$\bar{2}$] | [023] | [100] | [032] | [0$\bar{2}$3] | [100] |
| Σ13(510) | 21.8 | [05$\bar{1}$] | [015] | [100] | [051] | [0$\bar{1}$5] | [100] |
| Σ29(520) | 43.60 | [05$\bar{2}$] | [025] | [100] | [052] | [0$\bar{2}$5] | [100] |
| Σ49(853) | 43.57 | [$\bar{11}$ $\bar{2}$ 13] | [5$\bar{8}$3] | [111] | [$\bar{2}$ $\bar{11}$ 3] | [8$\bar{5}$3] | [111] |
| Σ91(11 6 5) | 53.99 | [$\bar{16}$ $\bar{1}$ 17] | [6 $\bar{11}$ 5] | [111] | [$\bar{1}$ $\bar{16}$ 17] | [11 $\bar{6}$ $\bar{5}$] | [111] |

# 3. Results

## 3.1. Structural and energetic characteristics of GBs

We begin by examining the atomic structures and energetics of twelve representative GBs summarized in Table 1. Fig. 2 shows the relaxed atomic structures of these GBs in fcc Ni; the corresponding GB structures in the alloy systems are essentially identical, apart from local lattice distortions induced by Cr and Fe solutes. Apart from the Σ3 coherent twin boundary, which exhibits a highly ordered and densely packed atomic arrangement, all GBs display typical high-angle GB characteristics. Specifically, they are composed of periodically arranged structural units rather than isolated dislocation cores, and atoms in the GB vicinity exhibit reduced atomic packing density accompanied by noticeable excess volume.

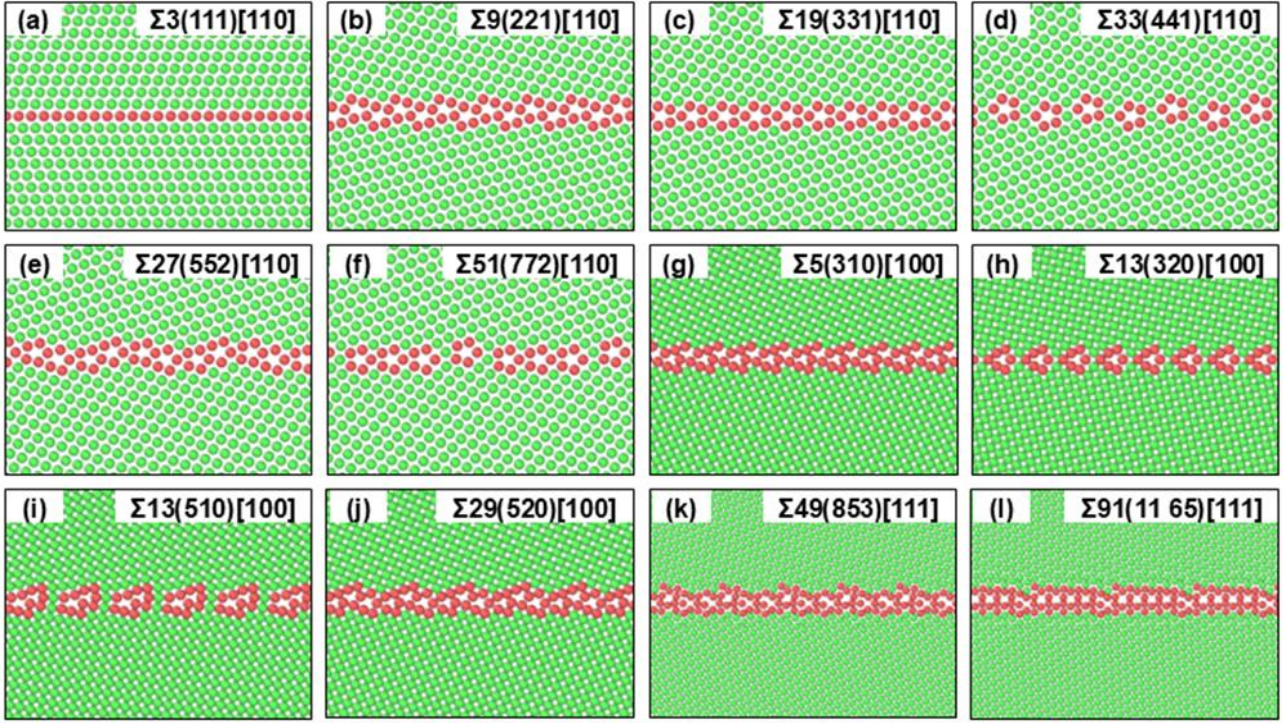

Figure 2. Atomic structures of the twelve representative GBs in fcc Ni. Green atoms denote atoms in perfect fcc coordination, whereas red atoms indicate defective sites with reduced local atomic packing. Except for the Σ3 coherent twin boundary, all GBs exhibit loosely packed structural units characteristic of high-angle GBs.

To assess the energetic stability of the constructed GBs, GB energies were calculated following:

$$E_{GB} = \frac{E_{tot} - N \cdot E_{coh}}{2A}, \quad (1)$$

where $E_{tot}$ is the total (potential) energy of the system, $E_{coh}$ is the cohesive energy of an atom, A is the area of the GB, and N is the number of atoms. GB energies for all twelve boundaries in both pure Ni and Alloy 690 are summarized in Table 2, along with comparison with previously reported values. The Σ3 twin boundary exhibits an exceptionally low GB energy, consistent with its compact structure, whereas the other GBs show relatively higher but comparable energies. The close agreement between the present results and literature data confirms the reliability of the employed interatomic potential and the validity of the constructed GB models. Overall, the pronounced differences in atomic packing density, excess volume, and energetics among GBs are expected to influence solute diffusion behavior and may therefore have important implications for subsequent grain boundary migration.

Table 2. Comparison of calculated GB energies with literature values (shown in brackets)[43], demonstrating the reliability of the employed interatomic potential.

| GB type | Ni | 690alloy |
|---|---|---|
| Σ3(111) | 0.06 (0.01) | 0.039 |
| Σ9(221) | 1.165 (1.122) | 1.22 |
| Σ19(331) | 1.089 (1.109) | 1.189 |
| Σ33(441) | 1.081 (1.024) | 1.116 |
| Σ27(552) | 1.22 (1.204) | 1.242 |
| Σ51(772) | 1.134 (1.112) | 1.173 |
| Σ5(310) | 1.284 (1.257) | 1.279 |
| Σ13(320) | 1.155 (1.174) | 1.11 |
| Σ13(510) | 1.325 (1.248) | 1.227 |
| Σ29(520) | 1.378 (1.359) | 1.39 |
| Σ49(853) | 1.375 (1.319) | 1.318 |
| Σ91(11 65) | 1.204 (1.221) | 1.202 |

## 3.2. GB migration under solute diffusion

To examine how these structural differences translate into GB mobility, GB migration behavior was systematically investigated using hybrid MD–SGCMC simulations. For each GB, simulations were performed for 10 ns, and the time interval from 0 to 7 ns—during which steady-state migration was observed—was selected for analysis. Representative GB displacement–time curves for the Type-690 system are shown in Fig. 3(a), where significant variations in migration behavior are observed among different GBs. Over the 7 ns simulation period, GB migration distances range from nearly zero to approximately 77 Å. Among the GBs examined, the Σ49(853) boundary exhibits the largest migration distance and the highest migration rate, followed by Σ29(520), Σ5(310), Σ91(11 65), and Σ27(552). In contrast, GBs such as Σ13(320) and Σ19(331) show almost no detectable migration.

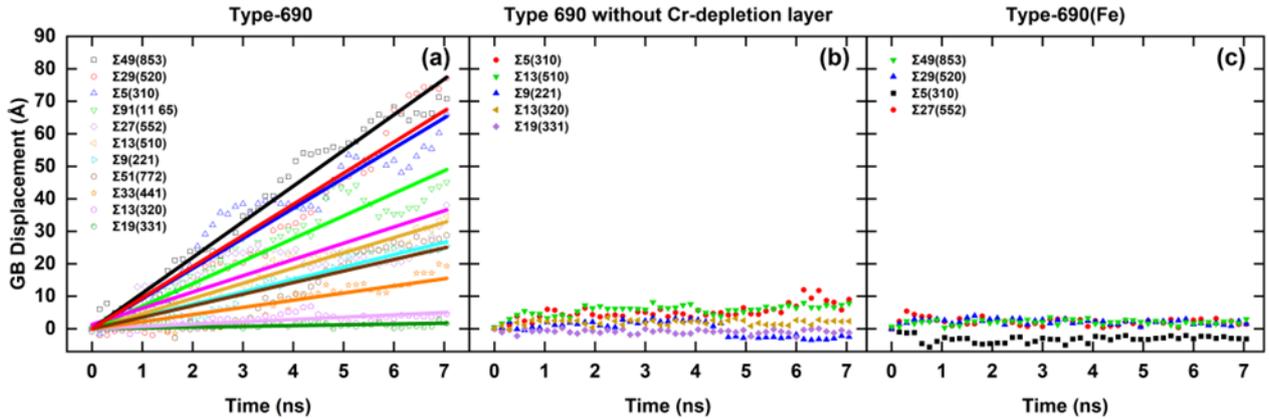

Figure 3. GB displacement as a function of simulation time for (a) Type-690, (b) Type-690 without MC exchanges, and (c) Type-690(Fe). Solid lines in (a) are linear fits for corresponding MD data. Directional and sustained GB migration is observed only in Type-690 under coupled diffusion and segregation, whereas the absence of MC operations results in negligible displacement dominated by thermal fluctuations.

Since a relatively high temperature of 1400 K was employed to accelerate diffusion within MD timescale, it is necessary to distinguish diffusion-induced GB migration from thermally activated interfacial random walk. To this end, control simulations were performed under identical conditions but with the SGCMC operations disabled (i.e., without solute-depleted layer). The resulting GB displacement curves are shown in Fig. 3b. In the absence of SGCMC, nearly all GBs exhibit negligible migration, with displacements generally less than 10 Å. These small fluctuations can be attributed to thermally driven random walk of the GBs. In comparison, the GB migration observed in Fig. 3(a) proceeds at significantly higher velocities and exhibits a persistent, unidirectional character once initiated. This behavior is characteristic of DIGM and is consistent with experimental observations reported in the literature [3,8,19,48], indicating the presence of a non-thermal driving force for GB migration.

To further elucidate the role of solute species, additional simulations were performed for selected GBs in the Type-690(Fe) system, as shown in Fig. 3(c). Notably, GBs such as Σ49(853) and Σ29(520), which migrate rapidly in the Type-690 system with Cr-depleted layer, exhibit negligible migration when Cr depletion is replaced by Fe depletion. The strong dependence of GB migration on both solute species and GB type implies that migration behavior cannot be explained solely by the presence of solute diffusion, motivating a detailed examination of solute redistribution and depletion near the GBs.

## 3.3. Correlation between GB migration and solute depletion

Numerous experimental studies have shown that GB migration corresponds closely to the formation of Cr depletion zones[8, 20, 25-28, 44-46] . To examine whether this correlation holds at the atomic scale in the present simulations, we analyzed the spatial distributions of solute atoms in the vicinity of representative GBs together with their migration behavior.

Figs. 4(a), (b), and (c) show the GB positions and elemental concentration profiles after 7 ns of simulation in the Type-690 system. Three GBs exhibiting distinct migration behaviors (Σ3(111), Σ19(331), and Σ49(853)) were selected for comparison. The Σ3(111) coherent twin boundary exhibits negligible GB migration, accompanied by minimal solute diffusion and no discernible Cr-depleted zone. This behavior is consistent with its compact atomic structure and low GB energy. Interestingly, the Σ19(331) GB develops a clear but very narrow Cr-depleted region, while exhibiting almost no GB migration. This observation indicates that although Cr diffusion and solute depletion occur along the Σ19(331) GB, the formation of a solute-depleted zone alone is insufficient to induce GB migration. In contrast, the Σ49(853) GB shows the largest migration distance among the examined boundaries and is associated with a pronounced Ni-enriched and Cr-depleted region. The Ni enrichment reaches approximately 10 at.%, corresponding to an equivalent depletion of Cr.

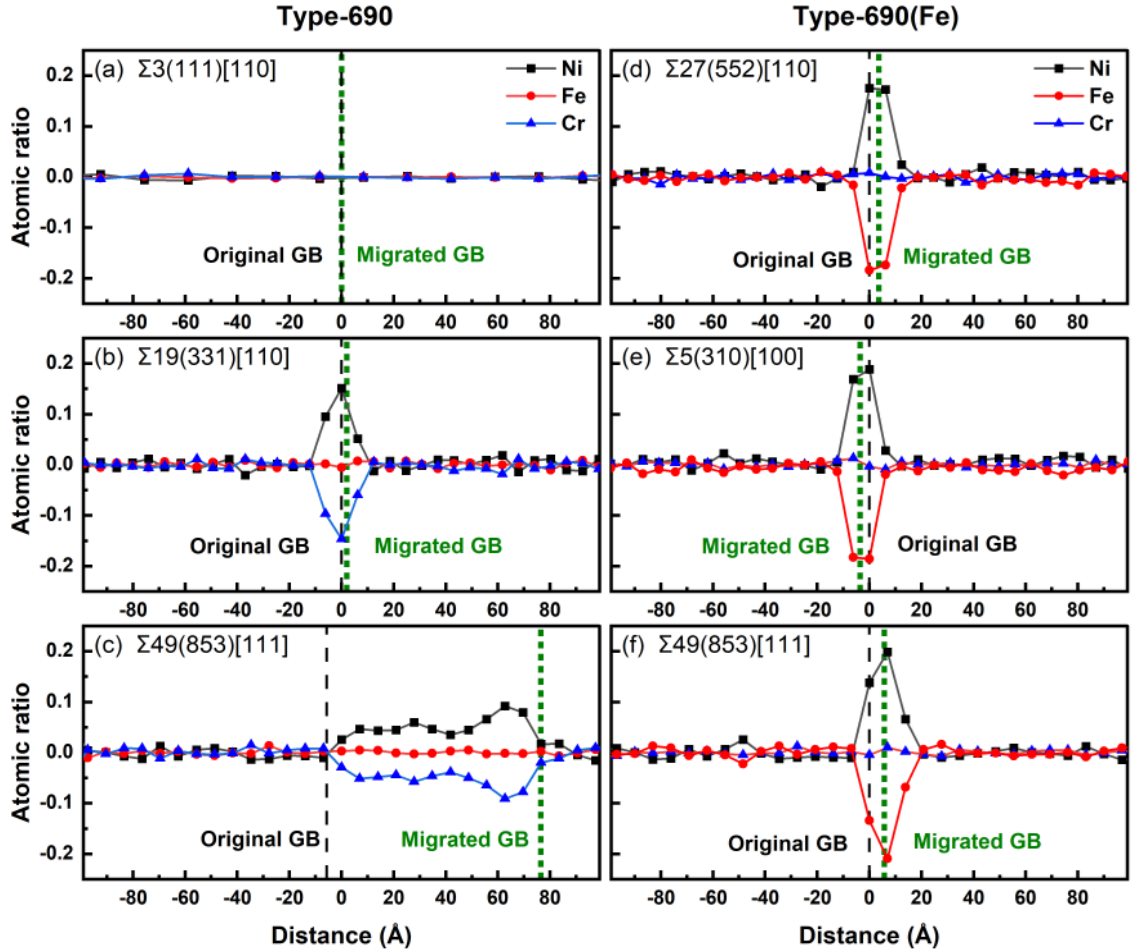

Figure 4. Change in solute concentration profiles normal to representative GBs after 7 ns of simulation. Panels (a), (b), and (c) correspond to Type-690 (Cr), whereas panels (d), (e), and (f) correspond to Type-690(Fe). In these simulations, pronounced GB migration is consistently associated with wide Cr depletion zones, whereas Fe depletion zones do not lead to sustained GB migration.

Overall, these results show remarkable agreement with experimental observations reported by Kuang *et al.* [8]. Specifically, random high-angle boundaries exhibit pronounced GB migration accompanied by wide Cr-depleted zones, consistent with the behavior observed for representative GBs in Fig. 4c. Incoherent twin boundaries develop narrow Cr-depleted regions without apparent migration, similar to the results shown in Fig. 4b, whereas coherent twin boundaries display negligible Cr depletion and essentially no GB migration, as illustrated in Fig. 4a.

To further assess whether the observed correlation between solute depletion and GB migration is specific to Cr, analogous analyses were performed for the Type-690(Fe) system, in which Fe serves as the primary diffusing solute. Corresponding results are shown in Figs. 4(d), (e), and (f) for the same representative GBs. Owing to active solute diffusion along the GBs, all three boundaries develop noticeable Ni-enriched and Fe-depleted regions, indicating that solute depletion is also readily established under Fe diffusion. However, despite the formation of clear Fe depletion zones, the migration distances of all examined GBs remain below 5 Å and can be regarded as negligible. These results demonstrate that the presence of a solute-depleted zone alone is insufficient to induce GB migration. Instead, GB migration in Ni-based alloys is highly solute-specific: while Cr depletion promotes pronounced migration, Fe depletion does not, despite comparable diffusion along the GBs. This contrast highlights the unique role of Cr in diffusion-induced grain boundary migration.

# 4. Discussion

## 4.1. Role of solute diffusion in GB migration

Classical descriptions of DIGM posit that solute diffusion along GBs in the presence of a solute source or sink can generate a driving force for GB migration, and that GBs with higher diffusivity are therefore expected to develop wider migration zones. This framework implicitly assumes that solute diffusion alone is sufficient to induce GB migration. To evaluate this assumption, solute diffusion behavior and GB migration were systematically compared for different GBs in both the Type-690 and Type-690(Fe) systems. As shown in Fig. 5, substantial variations in solute diffusivity are observed among different GBs. In the Type-690 system, several high-angle GBs exhibit fast Cr diffusion, whereas the Σ3 coherent twin boundary shows negligible diffusivity. While in the Type-690(Fe) system, we only examined representative GBs with relatively high Fe diffusivity.

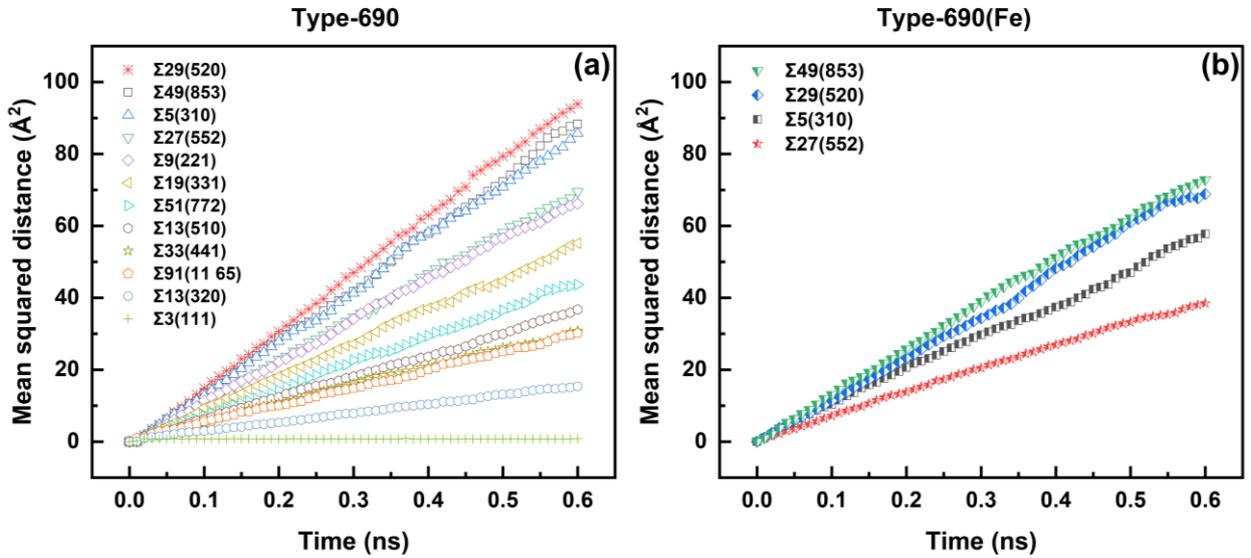

Figure 5. Mean squared displacement (MSD) of solute atoms along grain boundaries at 1400 K for (a) the Type-690 system and (b) the Type-690(Fe) system. The MSD was calculated for Cr in Type-690 and Fe in Type-690(Fe) within ±15 Å of the GB plane and along the GB direction.

By extracting effective diffusion coefficients from the linear regime of the MSD–time curves, we further correlate GB migration velocities with the corresponding solute diffusion coefficients (Fig. 6). While a general positive correlation can be identified as expected in the DIGM framework (i.e., GBs with higher diffusivity tend to exhibit higher migration rates), multiple notable deviations from this trend are observed. In the Type-690 system, the Σ19(331) GB exhibits a relatively high Cr diffusion coefficient, yet it shows negligible GB migration. In contrast, the Σ33(441) GB, which has a lower diffusion coefficient, displays substantial migration. Moreover, Σ33(441) and Σ91(11 6 5) exhibit nearly identical diffusion coefficients, yet their migration rates differ significantly, with Σ91(11 6 5) lying far outside the linear-fit trend. More strikingly, in the Type-690(Fe) system, the four GBs examined show substantial Fe diffusion, yet it does not lead to appreciable GB migration. These results demonstrate that solute diffusion rate alone is insufficient to account for the observed grain boundary migration behavior, indicating that an additional energetic factor beyond diffusion must be considered."

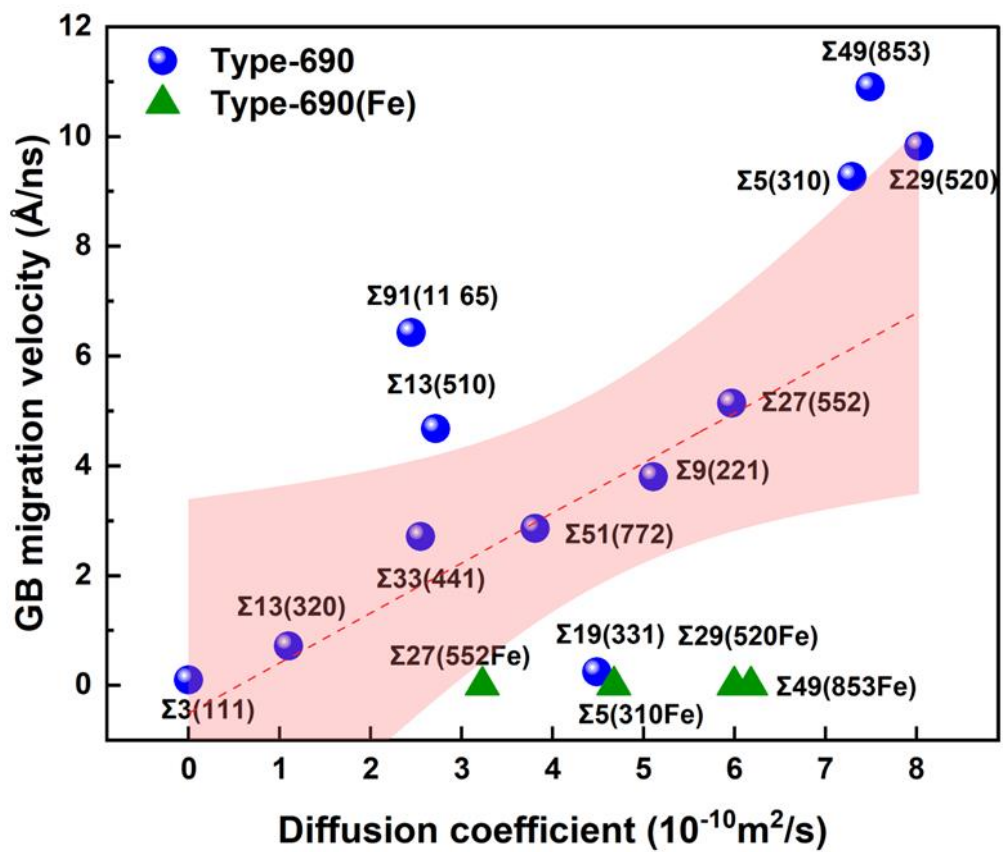

Figure 6. DIGM velocity as a function of the GB diffusion coefficient. Blue circles denote GBs in Type-690 with Cr diffusion, while green triangles denote GBs in Type-690(Fe) with Fe diffusion. The red line shows a linear fitting to the data, with the shaded area represents the 95% confidence interval. While an overall positive correlation is observed, pronounced deviations for specific GBs are evident.

## 4.2. Energetic origin of solute-induced GB migration

Motivated by the above observations that solute diffusion and depletion alone are insufficient to explain DIGM, we next quantify the energetic interaction between solute depletion zones and GBs. To this end, a solute-depleted region with a width of 30 Å was artificially introduced adjacent to the GB. For consistency across all simulations, the Cr/Fe concentration within the depleted zone was reduced from 30 at.% to 20 at.%, with the deficit compensated by Ni, matching the solute depletion levels observed in Fig. 4. The interaction energy was evaluated by systematically shifting the position of the solute-depleted zone relative to the GB plane. The resulting energy variations are shown in Fig. 7.

For the Type-690 system, the total energy reaches a maximum when the Cr-depleted zone coincides with the GB. This behavior indicates an intrinsic energetic incompatibility between the GB and the Cr-depleted region, such that the system can lower its energy by separating the GB from the depleted zone. This repulsive interaction provides a direct thermodynamic driving force for GB migration away from the Cr-depleted region. Notably, the magnitude of the migration energy barrier varies significantly among different GBs, indicating a strong dependence on GB character. For example, the Σ19(331) GB exhibits a very small barrier, whereas the Σ49(853) GB shows a substantially higher barrier, consistent with their contrasting migration behaviors observed in Fig. 4. Similarly, although Σ33(441) and Σ91(11 6 5) exhibit comparable diffusion coefficients (Fig. 5), their markedly different migration rates can be rationalized by their distinct interaction energy profiles.

In contrast, for the Type-690(Fe) system, the total energy reaches a minimum when the Fe-depleted zone is located at the GB. This indicates that the GB–depleted-zone configuration is energetically stable, and no driving force exists for GB migration. Consequently, even though Fe diffusion readily produces solute-depleted regions, GB migration is energetically unfavorable, explaining the negligible DIGM observed in Figs. 4 and 5. To further validate this interpretation, benchmark DFT calculations were performed. As listed in Table 3, the calculated segregation energy of Cr at the representative Σ5(310) GB is negative, whereas that of Fe is positive, in good agreement with the trends obtained from the MD simulations in Figure 7.

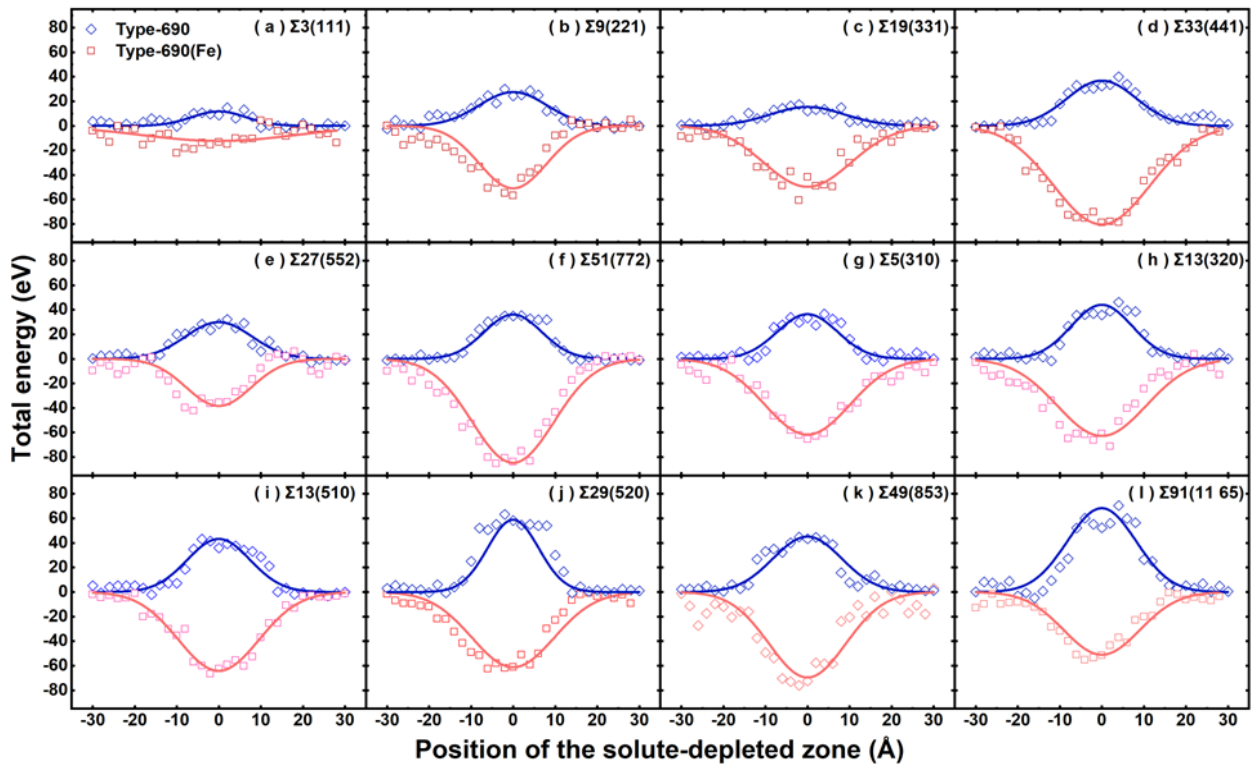

Figure 7. Interaction energy between solute-depletion zones and representative GBs in Type-690 and Type-690(Fe) systems, as a function of their relative position. The depleted zones correspond to Cr in

Type-690 and Fe in Type-690(Fe), with the solute concentration reduced from 30 at.% to 20 at.% and the deficit compensated by Ni. Blue symbols denote Cr depletion zones in Type-690, red symbols denote Fe depletion zones in Type-690(Fe), and solid curves are Gaussian fitting for corresponding data.

Table 3. Comparison of segregation energies (unit in eV) of Cr and Fe in Ni at the Σ5(310) GB obtained from DFT and MD calculations. The consistent segregation trends between the two methods validate the reliability of the interatomic potential used in the MD simulations.

| Solute type | MD | DFT |
|---|---|---|
| Cr | -0.353 | -0.320 |
| Fe | 0.156 | 0.123 |

These results reveal three distinct solute–GB interaction regimes in Ni-based alloys, leading to fundamentally different GB migration behaviors, as schematically illustrated in Fig. 8. When solute diffusion along the GB is slow, solute depletion does not develop, and the GB remains stable without observable migration (Fig. 8a). When solute diffusion along the GB is sufficiently fast, a solute-depleted zone can form locally along the GB. However, in the case of weak solute segregation or anti-segregation (such as for Fe or for Cr at certain GBs), the depleted zone does not extend laterally. The coincidence of a solute-depleted region with the GB is then energetically favorable, stabilizing the boundary position and suppressing sustained GB migration, even though limited displacements may occur due to thermal fluctuations (Fig. 8b). In contrast, for solutes exhibiting strong positive segregation to GBs, such as Cr, fast GB diffusion produces localized Cr depletion that is energetically incompatible with the segregating GB. This generates a chemical driving force that pulls the GB toward the solute-rich matrix. As the migrating GB encounters solute-rich material, rapid solute diffusion replenishes segregation at the boundary while simultaneously extending the depleted zone. This coupled diffusion–segregation–migration process sustains persistent and directional DIGM accompanied by a solute-depleted wake, as shown in Fig. 8c.

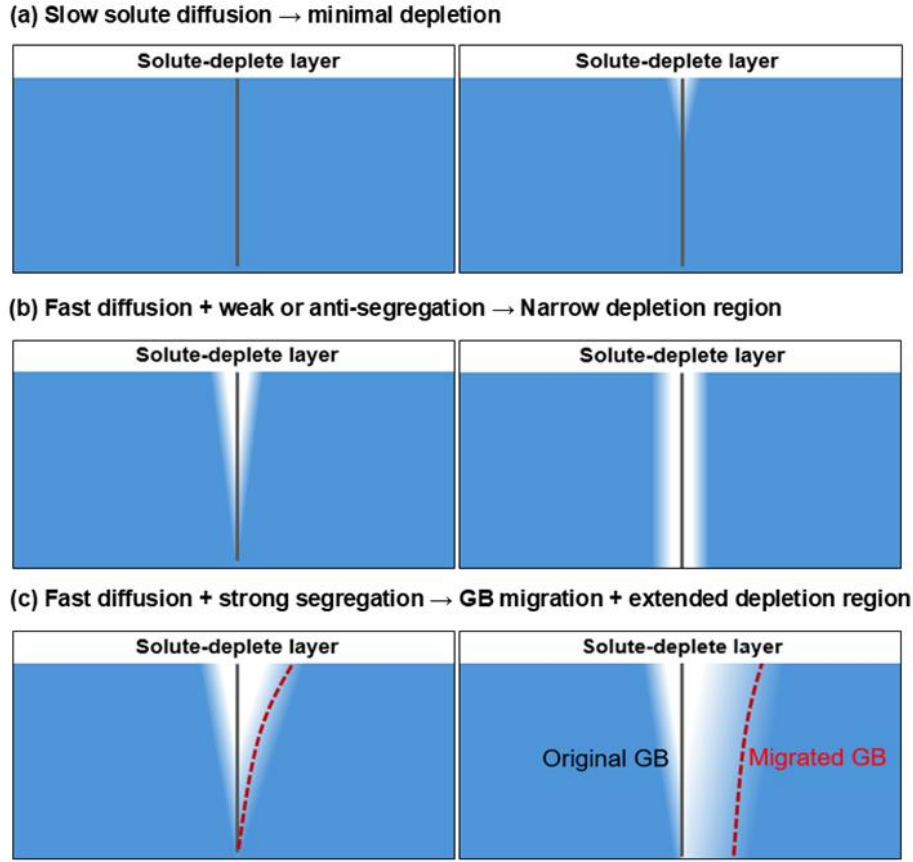

Figure 8. Schematic illustration of GB migration behavior under surface-driven solute depletion for different diffusion and solute–GB interaction regimes. (a) Slow solute diffusion along the GB, resulting in negligible solute depletion and no GB migration. (b) Fast solute diffusion leading to the formation of a localized solute-depleted zone along the GB; however, due to weak or anti-segregation solute–GB interactions, the depletion zone does not extend laterally and does not induce sustained GB migration. (c) Fast solute diffusion combined with strong positive solute segregation at the GB, resulting in lateral expansion of the solute-depleted zone and sustained GB migration.

## 4.3. Coupled effects of segregation and diffusion on grain boundary migration

The results in Section 4.2 establish that solute depletion zones interact energetically with GBs in a solute-specific manner. Building on this energy landscape, we now integrate diffusion and segregation effects to formulate a unified description of DIGM kinetics across different GBs.

For a given GB, the migration velocity is governed not solely by how fast solute atoms diffuse along the boundary, but also by how strongly the resulting solute-depleted zone interacts with the GB. This can be quantified by incorporating both kinetic and thermodynamic contributions:

$$v = M \cdot D \cdot \max(E, 0), \qquad (2)$$

Where GB migration velocity, $v$, is quantified by the GB diffusion coefficient $D$ and the normalized segregation strength $E$ (energy change per unit area associated with the solute-depleted zone in Fig. 7), scaled by a coefficient $M$. As a result, GB migration can be described by the combined parameter $D \times E$, which incorporates both kinetic and thermodynamic contributions. The resulting correlation between the DIGM velocity $v$ and $D \times E$ is shown in Fig. 9, where a clear linear relationship is observed for all GBs in both Type-690 and Type-690(Fe) systems.

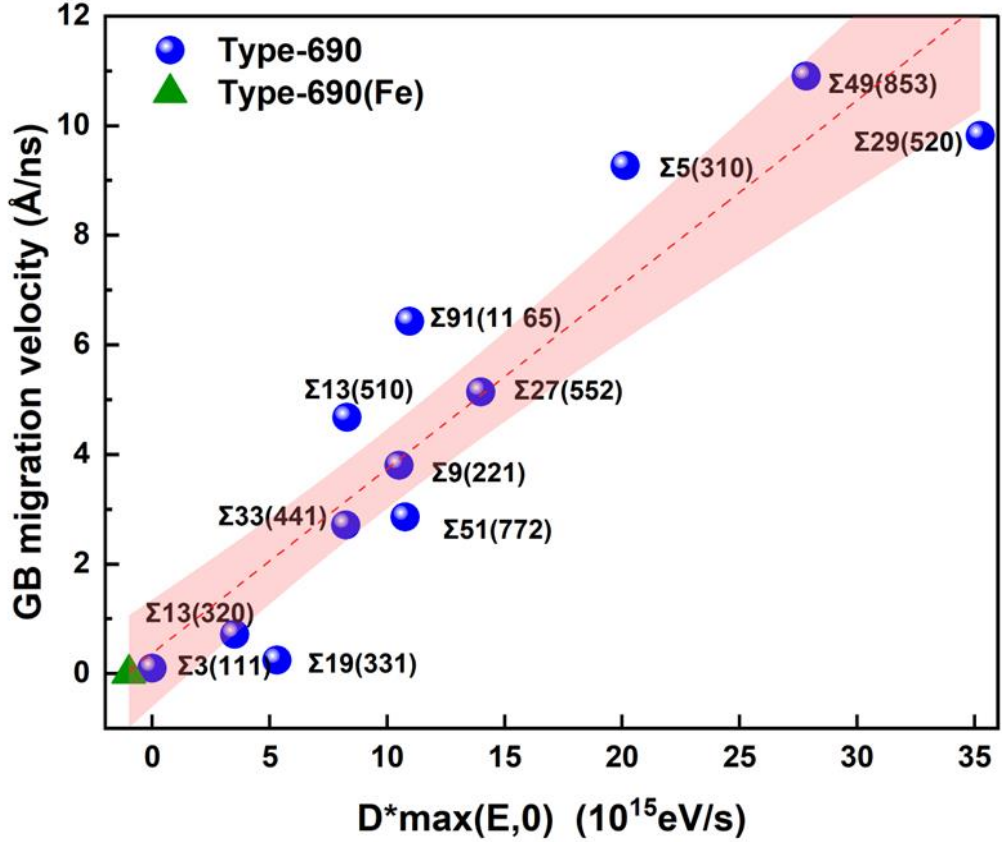

Figure 9. DIGM velocity as a function of the combined parameter $D \times E$ for different GBs. Blue circles denote Type-690 with Cr diffusion, and green triangles denote Type-690(Fe) with Fe diffusion (slightly offset for clarity, as the corresponding data points cluster near the origin). The linear relationship (red line) highlights the coupled roles of solute diffusion and segregation energetics in governing GB migration, and the shaded area represents the 95% confidence interval.

This quantitative framework naturally explains several seemingly anomalous features previously observed in Fig. 5. For instance, the Σ19(331) GB exhibits strong Cr diffusion yet shows negligible migration because its interaction energy $E$ is extremely small. Conversely, Σ33(441) and Σ91(11 65) GBs possess similar diffusion coefficients but markedly different migration velocities due to their distinct interaction energies. In the Type-690(Fe) system, the interaction energy $E$ is negative, which stabilizes the GB position and suppresses DIGM regardless of the diffusion rate.

An important and initially non-intuitive implication of this framework is that strong solute segregation can promote, rather than suppress, solute-depletion-driven GB migration. In the case of Cr, positive segregation lowers the GB energy only when a sufficiently high Cr concentration can be maintained at the boundary. Under oxidizing conditions, however, rapid GB diffusion continuously removes Cr from the boundary to sustain surface chromia formation, leading to the development of a Cr-depleted zone that coincides with the GB. Once this occurs, the energetic stabilization associated with Cr segregation can no longer be sustained, resulting in an energy penalty that drives the GB away from the depleted region and toward the Cr-rich matrix. Consequently, segregation does not prevent depletion; instead, it amplifies the energetic driving force once depletion develops, thereby promoting sustained DIGM. This mechanism, referred to here as segregation-induced depletion-driven migration, is fully consistent with the interaction energies quantified in Section 4.2.

In the context of inter-granular SCC and DIGM, the role of GB diffusion in corrosion resistance is inherently mechanism-dependent, governed by the coupling between solute diffusion and segregation energetics. Although solute diffusion is required for redistribution during oxidation, excessively fast diffusion along GBs can continuously remove solute from the boundary and promote the formation of localized depletion zones. The present results demonstrate that GB diffusion alone cannot be used to assess corrosion resistance. Instead, resistance to localized degradation is favored when solute redistribution does not lead to persistent depletion, which occurs when GB diffusivity is limited or segregation interactions are weak. This framework clarifies the apparently diverse observations reported in the literature by showing that the effect of diffusion critically depends on its coupling with segregation and GB energetics. More broadly, these results highlight that accurate prediction of GB stability in service environments requires consideration of coupled diffusion–segregation effects involving multiple solute species, rather than diffusion kinetics alone. Such insights provide a mechanistic basis for microstructural and compositional design strategies aimed at mitigating DIGM and enhancing long-term SCC resistance.

# 5. Conclusions

In this work, diffusion-induced grain boundary migration (DIGM) in Alloy 690 was systematically investigated using hybrid molecular dynamics and semi-grand canonical Monte Carlo simulations, with explicit consideration of coupled solute diffusion and segregation effects across a broad range of grain boundary (GB) characters. The main conclusions can be summarized as follows:

1. DIGM in Alloy 690 is strongly solute-dependent. Cr diffusion leads to pronounced GB migration accompanied by Cr depletion, whereas Fe diffusion—despite forming local solute depletion zones—does not induce sustained GB migration, demonstrating that diffusion alone is insufficient to drive DIGM.
2. The driving force for DIGM arises from the energetic interaction between solute-depleted zones and GBs, which is governed by solute segregation behavior. Positive segregation of Cr renders the coincidence of a Cr-depleted zone with the GB energetically unfavorable, driving the GB toward the Cr-rich matrix. In contrast, Fe exhibits anti-segregation behavior, which stabilizes the GB at Fe-depleted regions and suppresses migration. As a result, strong solute segregation—often regarded as stabilizing GBs—can promote depletion-driven GB migration once depletion develops.
3. GB migration kinetics are governed by the coupled effects of solute diffusivity and segregation strength. A unified parameter incorporating both quantities quantitatively captures GB-dependent DIGM velocities across different GB characters and solute systems.

Overall, these results establish a unified mechanistic framework linking solute diffusion, segregation energetics, and GB migration in Cr-containing alloys. The present findings provide mechanistic insight into experimentally observed correlations between GB character, Cr depletion, and intergranular degradation, and offer guidance for mitigating DIGM and associated degradation through compositional and microstructural control.

**Acknowledgement:**

This work was financially supported by National Natural Science Foundation of China (No.: 52401010; 12375260), Hunan Provincial Natural Science Foundation of China (No.: 2025JJ40005). State Key Laboratory of Cemented Carbide Construction Project (No.: 2024ZYT006). We acknowledge Hefei Advanced Computing Center for providing computing resources.

**Competing interests**

The authors declare no competing interests.

**Data availability**

The data generated and/or analyzed within the current study will be made available upon reasonable request to the authors.

**Author contributions**

**Yuntian Jiang:** Methodology, Formal analysis, Investigation, Data Curation, Writing - Original Draft, Visualization. **Shuai Zhang:** Writing - Review & Editing. **Zhuoming Xie:** Resources, Writing - review & editing. **Xiaohan Bie:** Writing - review & editing. **Huiqiu Deng:** Writing - review & editing. **Wangyu Hu:** Resources, Supervision, Writing - review & editing. **Jie Hou:** Conceptualization, Methodology, Validation, Formal analysis, Resources, Data curation, Writing - Review & Editing, Supervision, Project administration, Funding acquisition